\documentclass[a4paper,fleqn,usenatbib,useAMS]{mnras}

\newcommand{\beqa}{\begin{eqnarray}}
\newcommand{\eeqa}{\end{eqnarray}}

\usepackage{graphicx}
\usepackage{epsfig}

\usepackage{graphicx}
\usepackage{dcolumn}
\usepackage{bm}%

\usepackage{graphicx}	
\usepackage{amsmath}	
\usepackage{amssymb}	
\usepackage{multicol}        
\usepackage{bm}		
\usepackage{pdflscape}	

\title[MNRAS \LaTeX\ guide for authors]{Constraints on the 
small scale curvature perturbation using Planck-2015 data}

\author[Yupeng Yang]{Yupeng Yang$^{1}$$^{2}$$^{3}$\thanks{Contact e-mail: \href{mailto:ypyang@aliyun.com}{ypyang@aliyun.com}}
\\
$^{1}$School of Physics and Physical Engineering, Qufu Normal University, Qufu, Shandong, 273165, China\\
$^{2}$Collage of Physics and Electrical Engineering, Anyang Normal University, Anyang, 455000, China\\
$^{3}$Joint Center for Particle, Nuclear Physics and Cosmology, Nanjing, 210093, China}

\pubyear{2015}

\begin{document}
\label{firstpage}
\pagerange{\pageref{firstpage}--\pageref{lastpage}}
\maketitle

\begin{abstract}
The particles emitted from PBHs through the Hawking radiation 
have interactions with the particles present in the Universe. 
Due to the interactions, the evolution of the intergalactic medium (IGM) is changed 
and the changes have imprints on the 
anisotropies of the cosmic microwave background (CMB). 
In this paper, we focus on the PBHs with the lifetime in the range of 
$10^{13}s \lesssim \tau_{\rm PBH} \lesssim 10^{17}s$, 
corresponding to the mass range of $2.8\times 10^{13}\mathrm{g} \lesssim M_{\mathrm{PBH}} 
\lesssim 2.5\times 10^{14}\mathrm{g}$. 
We update the constraints on the initial mass fraction of PBHs 
using the Plank-2015 data. We find that the optimistic upper limits are 
$4\times 10^{-29}\lesssim \beta(M_{\mathrm{PBH}}) \lesssim 5\times 10^{-28}$, depending on the mass of PBH. 
The formation of PBHs is related to the primordial 
curvature perturbations. Therefore, using the constraints 
on the initial mass fraction of PBHs, we get the upper limits on the 
power spectrum of primordial curvature perturbation. For the investigated mass range of PBHs, 
corresponding to the range of scales 
$8.9 \times 10^{15}\ \mathrm{Mpc^{-1}}\lesssim k \lesssim 2.8 \times 10^{16}\  \mathrm{Mpc^{-1}}$, 
we find that the upper limits change slightly with a value of $\mathcal{P}_\mathcal{R}(k) \sim 0.0045$, 
and the limits are slightly stronger compared with the previous results.

\end{abstract}

\maketitle

\section{Introduction}

It has been predicted by many inflation models that the power spectrum of primordial 
curvature perturbation, $\mathcal{P}_\mathcal{R}(k)$, is nearly scale invariant~\citep{model_1}. 
On large scales, $10^{-4}\ \mathrm{Mpc^{-1}} 
\lesssim k \lesssim 1\ \mathrm{Mpc^{-1}}$, 
the constraint is $\mathcal{P}_\mathcal{R}(k) \sim 10^{-9}$, 
which is obtained mainly from the CMB, Lyman-$\alpha$ forest and large scale 
structures \citep{cmb_2,lyman,large}. 
On small scales, $1\ \mathrm{Mpc^{-1}}\lesssim k \lesssim 10^{20}  \ \mathrm{Mpc^{-1}}$, 
the upper limit is $\mathcal{P}_\mathcal{R}(k) \sim 10^{-2}$, 
which is obtained from the researching on the 
PBHs~\citep{Josan:2009,carr}. 
Recently, a new kind of dark matter structures named ultracompact dark matter 
minihalos (UCMHs) has been suggested, and it is found that researching on UCMHs 
give much more stringent limits, $\mathcal{P}_\mathcal{R}(k) \lesssim 10^{-5}$ 
for the range of scales $5\ \mathrm{Mpc^{-1}} \lesssim k\lesssim 10^{8} \ \mathrm{Mpc^{-1}}$
~\citep{josan,Bringmann_1,yyp_neutrino,fangdali,scott_2015}. 
A flaw of the limits obtained from UCMHs 
is that they mostly depend on the non-gravitational properties of dark matter (DM) particles. 
\citet{prl_1,prl_2} suggested that 
the energy deposited into the baryonic fluid due to the silk damping 
effect can also be used to constrain the primordial curvature perturbation, and 
they found a boost upper limit $\mathcal{P}_\mathcal{R}(k)\sim 0.06$ in the 
range of scales $10^4\ \mathrm{Mpc^{-1}}\lesssim k \lesssim 10^5 \ \mathrm{Mpc^{-1}}$.

PBHs can form in the early Universe via the collapse of large density perturbations. 
After formation, PBHs can emit photons, electrons, 
neutrinos, quarks and other particles through the Hawking radiation. 
The particles emitted from PBHs have interactions with the particles 
present in the Universe. As a result, the evolution 
of IGM is changed, and the changes have imprints on the CMB or the other astrophysical processes~\citep{mack_21}. 
Therefore, the data of CMB can be used to constrain the 
initial mass fraction of PBHs $\beta(M_{\mathrm{PBH}})$. On the other hand, the formation of PBHs is related to the primordial 
curvature perturbations. Using the constraints on the initial mass fraction of PBHs, one 
can get the limits on the primordial curvature perturbations. 
It has been pointed that the influences of PBHs on the evolution of IGM, 
which include heating and ionzing on IGM, are similar to that of DM decay~\citep{carr,mack_21}. 
Specifically, one can treat the lifetime 
of a PBH ($\tau_{\rm PBH}$) as the inverse of the DM decay rate 
($\Gamma_{\rm DM}^{-1}$), and we follow this method for our calculations. 
The influences of DM decay on the evolution of IGM have been investigated by previous works~\citep{xlc_decay,lz_decay,DM_2015}.  
~\citet{carr} have shown that 
the relevant results from the DM decay obtained using the WMAP3 data can be used to constrain 
the initial mass fraction of PBHs, and the constraints 
are stronger than that obtained using other observational data. 
In this paper, we update the constraints on the initial mass fraction of PBHs 
using the Planck-2015 data, then we use these constraints to 
get the limits on the power spectrum of primordial curvature perturbation. 
We find that for the optimistic efficiency of the energy deposited into IGM, $f=1$, the limits on 
$\beta(M_{\mathrm{PBH}})$ and 
$\mathcal{P}_\mathcal{R}(k)$ are about two and one order of magnitude stronger 
than that of previous results, respectively~\citep{Josan:2009,carr}.

This paper is organized as follows: In Sec. II, 
taking into account the influences of PBHs on the evolution of IGM, 
we get the constraints on the initial mass fraction of PBHs using the Planck-2015 data. 
In Sec. III, the limits on the power spectrum of primordial curvature perturbation 
are obtained using the relevant results given in Sec.II. 
The conclusions are given in Sec. IV.

\section{The influences of PBHs on the evolution of IGM and constrains on the PBHs initial mass fraction}
Through the Hawking radiation, PBHs can emit different particles 
depending on their 
masses~\citep{pbhs_emit_1,pbhs_emit_2,pbhs_emit_3}. 
The temperature and lifetime of a PBH are~\citep{pbhs_2016} 

\beqa
T_{\rm PBH} \approx 106\left(\frac{M_{\rm PBH}}{10^{14}\rm g}\right)^{-1}\ \rm MeV \\
\tau_{\rm PBH} \approx 2.7\times 10^{14}\left(\frac{M_{\rm PBH}}{10^{14}\rm g}\right)^3\frac{1}{f(M_{\rm PBH})}\ s,
\eeqa
where $f(M_{\rm PBH})$ is a function of PBH mass~\citep{pbhs_2016,carr,Josan:2009}. 
For the mass range $M_{\rm PBH} = 10^{15}{\rm g} - 10^{17}{\rm g}$\ ($T_{\rm PBH} = 0.106~\rm MeV-10.6~\rm MeV$), 
electrons can be emitted, and muons can be emitted for smaller masses 
$M_{\rm PBH} = 10^{14}{\rm g}-10^{15}{\rm g}$\ ($T_{\rm PBH} = 10.6~\rm MeV - 106~\rm MeV$). 
Pions can be emitted for PBHs in the mass range $M_{\rm PBH}\lesssim 5 \times 10^{14}\rm g$, 
where PBHs have completed their evaporation at the present epoch. 
For the temperature of PBHs exceeding the QCD confinement scales, 
$\Lambda_{\rm QCD} = 250 - 300~\rm MeV$\ ($M_{\rm PBHs} = 3.5\times 10^{13}{\rm g} - 4.2\times 10^{13}{\rm g}$), 
other fundamental particles such as quarks and gluons can be emitted~\citep{carr,pbhs_2016}. 
In this paper, we focus on the PBH with a lifetime in the range of $10^{13}s \lesssim \tau_{\rm PBH} \lesssim 10^{17}s$, 
which corresponds to the mass range of $10^{13}\ \rm g \lesssim M_\mathrm{PBHs} \lesssim 10^{14}\ \rm g$. 
For this mass range, PBHs complete their evaporation between the epoch of recombination ($z \sim 1100$) 
and reionization ($z \sim 6$). 

The particles emitted by PBHs have interactions with particles present in the Universe. 
The evolution of IGM can be changed due to the interactions. The main influences on IGM 
are ionization and heating, which have imprints on the CMB~\citep{1985apj,xlc_decay,mack_21}. 
Although the mechanism of Hawking radiation is not the same as that of DM decay, 
the influences of PBHs on the evolution of IGM are similar to the DM decay case~\citep{mack_21}.
\footnote{Here we do not consider the accretion of gas onto PBHs. 
The energy released from the accretion process can deposit into IGM, and the evolution 
of IGM is also changed. For this case, the PBHs is not similar to the DM decay case.} 
Specifically, for the purposes of our calculations, 
the lifetime of PBH with a fixed mass, $\tau_{\rm PBH}(M_{\rm PBH})$, 
can be equivalent to the inverse of the DM decay rate $\Gamma_{\rm DM}^{-1}$. 
Following the methods given by~\citet{xlc_decay,lz_decay}, 
taking into account the Hawking radiation of PBHs, 
the evolution of the ionization degree ($x_e$) and 
the temperature of IGM ($T_k$) can be written as

\beqa
(1+z)\frac{dx_{e}}{dz}=\frac{1}{H(z)}\left[R_{s}(z)-I_{s}(z)-I_{\rm PBH}(z)\right],
\eeqa

\beqa
(1+z)\frac{dT_{k}}{dz}=\frac{8\sigma_{T}a_{R}T^{4}_{\rm CMB}}{3m_{e}cH(z)}\frac{x_{e}}{1+f_{\rm He}+x_{e}}
(T_{k}-T_{\rm CMB})\\ \nonumber
-\frac{2}{3k_{B}H(z)}\frac{K_{\rm PBH}}{1+f_{\rm He}+x_{e}}+T_{k}, 
\eeqa
where $R_{s}(z)$ and $I_{s}(z)$ are the standard recombination rate and ionization rate, respectively. 
$I_{\rm PBH}$ and $K_{\rm PBH}$ are the ionization rate and heating rate caused by PBHs, 
which can be written as

\beqa
I_{\rm PBH} = \chi_{i}f^{'}(f\Omega_\mathrm{PBH}/\Omega_{b})(m_{b}c^2/E_{b})\tau_\mathrm{PBH}^{-1}\mathrm{e}^{-t\tau_\mathrm{PBH}^{-1}},
\label{ion}
\eeqa

\beqa
K_{\rm PBH} = \chi_{h}f^{'}(f\Omega_\mathrm{PBH}/\Omega_{b})m_{b}c^2\tau_\mathrm{PBH}^{-1}\mathrm{e}^{-t\tau_\mathrm{PBH}^{-1}},
\label{heat}
\eeqa
where $f$ is the fraction of the energies deposited in the IGM 
and it is generally a function of redshift~\citep{energy_function}. 
In this paper, we treat $f$ as a free parameter and 
the relevant discussions are given in following sections. 
$f^{'}$ is the fraction of electrons and positrons among 
the particles emitted by PBHs. 
As mentioned above, PBHs can emit different particles 
depending on their masses. Moreover, the mass of PBH can be changed with 
the particle emission. 
For the masses of PBHs considered in this paper, the most influences of particles 
on the IGM are caused by electrons, positrons and photons~\citep{xlc_decay}. 
For the photons, due to the processes of pair production, 
the energy deposited in the IGM 
can be treated as that of the electrons and positrons~\citep{mack_21}. 
Therefore, for the masses of PBHs considered in this paper, following ~\citet{carr}, we adopt $f^{'} = 0.1$. 
$\Omega_{\rm PBH}$ and $\Omega_{b}$ are the density parameters 
of PBH and baryon, respectively. 
$E_{b} = 13.6\ \rm eV$ is the ionization energy. 
$\chi_{i}$ and $\chi_{h}$ are the fractions of deposited energy 
for the ionization and heating of IGM,  
and which have been computed in detail by~\citet{1985apj}. In this paper, we use 
the forms suggested by~\citet{xlc_decay}, $\chi_{i} = (1-x_e)/3$, 
$\chi_{h} = (1+2x_e)/3$. 
The more accurate calculations and discussions about $\chi_{i,h}$ have been done 
by~\citet{chi_1,chi_2}. According to the discussions given by~\citet{chi_1}, 
the forms of $\chi_{i,h}$ used in this paper are enough for our 
calculations and it is excepted that the much more accurate ones can 
effect our final results slightly.\footnote{For more detailed 
discussions, one can refer to the Sec.V given by~\citet{chi_1}.}

In order to investigate the evolution of IGM described by Eqs.~(\ref{ion}) and~(\ref{heat}), 
we have modified the public code RECFAST~\footnote{http://camb.info} 
to account for the contributions of PBHs. 
For parameter fitting, we have used the Markov
Chain Monte Carlo (MCMC) techniques. We modify the 
public MCMC code CosmoMC\footnote{http://cosmologist.info/cosmomc/} in order to vary the new parameters along with 
the cosmological parameters. As shown above, for our purposes, 
we should consider the set of six cosmological parameters, 
\{$\Omega_{b}h^2, \Omega_{c}h^2,\theta, \tau, n_s, A_s$\}, and 
two new parameters, $\tau_\mathrm{PBH}^{-1}$ and 
$\zeta \equiv f\Omega_\mathrm{PBH}$. $\Omega_{b}h^2$ and 
$\Omega_{c}h^2$ are the density parameters of baryon and dark matter, $\theta$ is the
ratio of the sound horizon at recombination to its angular diameter 
distance multiplied by 100, $\tau$ is the optical
depth, $n_s$ and $A_s$ are the the spectral index and
amplitude of the primordial density perturbation power
spectrum. For the PBH with a lifetime larger than the age of the Universe, 
$\tau_\mathrm{PBH} \gtrsim 10^{17}s$, the factor 
$\mathrm{e}^{-t\tau_\mathrm{PBH}^{-1}}$ in Eqs.~(\ref{ion}) and~(\ref{heat}) is close to 1. 
Therefore, only one new combined 
parameter $\tau_\mathrm{PBHs}^{-1}\zeta \equiv \tau_\mathrm{PBHs}^{-1}f\Omega_{\rm PBH}$ 
is needed to be fitted~\citep{lz_decay,xlc_decay,mack_21,DM_2015}. However, for the PBH with a short 
lifetime as being considered in this paper, 
one must fit two new parameters $\tau_\mathrm{PBHs}^{-1}$ and 
$\zeta \equiv f\Omega_\mathrm{PBH}$ simultaneously~\citep{lz_decay,xlc_decay}. 

\begin{figure}
\epsfig{file=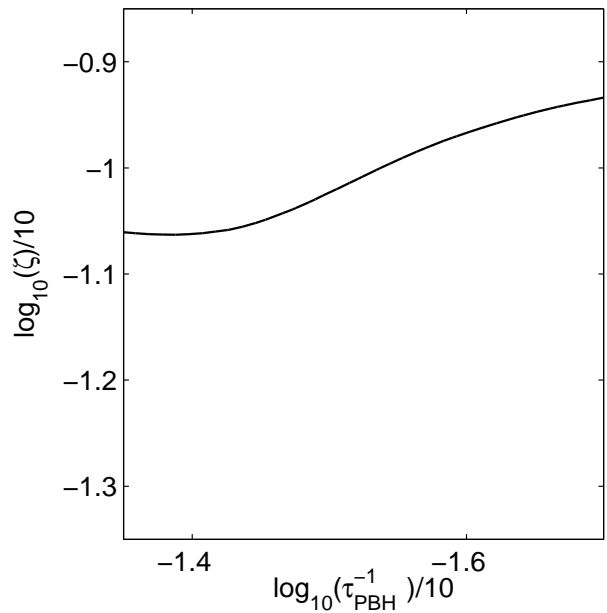,width=0.45\textwidth}
\caption{The constraints (95\% C.L.) on the PBH parameters in the two-dimensional parameter 
space [~$\rm log_{10}(\tau^{-1}_{\rm PBHs})/10$, 
$\rm log_{10}(\zeta)/10$~], which are obtained after 
marginalization over the other parameters (6 cosmological parameters). 
The region above the black line is excluded by the Planck-2015 data. 
In this plot, we have set the lifetime range 
of PBH as $10^{13}s \leq \tau_{\rm PBH} \le 10^{17}s$.}
\label{fig:2d}
\end{figure}

For our purposes, for the final constraints on the power spectrum of curvature perturbation, 
we are most interested in the two new parameters 
$\tau_\mathrm{PBH}^{-1}$ and $\zeta$. After running the MCMC code CosmoMC with the Planck-2015 data, 
the constraints on the parameters are obtained. 
For our purpose, in Fig.~\ref{fig:2d}, we plot the constraints on the PBH parameters 
in the two-dimensional parameter space [~$\rm log_{10}(\tau^{-1}_{\rm PBHs})/10$, 
$\rm log_{10}(\zeta)/10$~], which are obtained after 
marginalization over the other parameters (6 cosmological parameters), 
and the constraints on the other parameters are not shown. 
As shown in Fig.~\ref{fig:2d}, the region above the black line is excluded by the Planck-2015 data 
and the allowed region is under the black line. 
The boundary line can be written in a simple form and the allowed 
parameters space is given by

\beqa
\frac{1}{10}\mathrm{log}_{10}(\zeta) < p_{1}+p_{2}x+p_{3}x^2+p_{4}x^3+p_{5}x^4+p_{6}x^5, 
\label{eq2}
\eeqa
where $p_{1} = -555.461, p_{2} = -1836.23, p_{3} = -2418.65, 
p_{4} = -1583.48, p_{5} = -515.301, p_{6} = -66.6964$ and 
$x = \frac{1}{10}\mathrm{log}_{10}(\tau_\mathrm{PBH}^{-1})$. 

The formation of PBHs is related to the large density perturbations 
existed in the early Universe. 
One of the most important issues is the initial mass fraction of PBHs, 
which is defined as 
$\beta \left(\rm M_{PBH}\right) \equiv \rho^{i}_\mathrm{PBH}/\rho^{i}_\mathrm{crit}$~\citep{carr}, 
and it stands for the fraction of the horizon mass which collapses into the formation of PBHs. 
$\rho^{i}_\mathrm{crit}$ is the critical energy density at the formation time of PBHs. 
The initial mass faction has a relation to 
the parameter $\zeta$ as~\citep{Josan:2009, carr} 


\beqa
\beta \left(M_{\rm PBH}\right) =  1.5 \times 10^{-18} ~\zeta \left(\frac{M_{\rm PBH}}{5\times10^{14}\rm g}\right)^{1/2}.
\label{eq:beta}
\eeqa

\begin{figure}
\epsfig{file=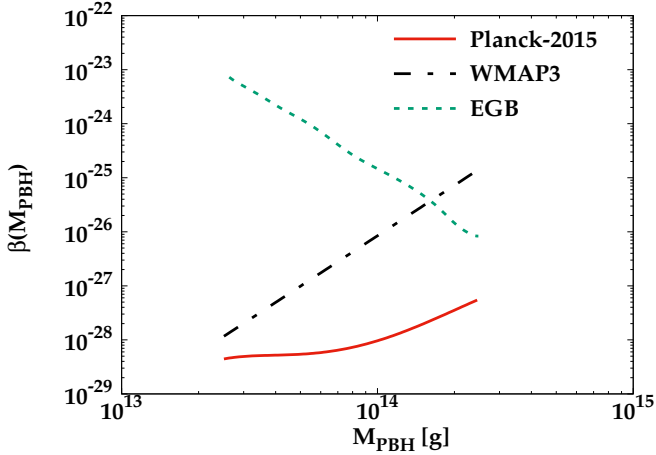,width=0.5\textwidth}
\caption{The upper limits (95\% C.L.) on the initial mass fraction of PBH in the 
mass range of $2.8 \times 10^{13} \rm g \lesssim M_{\rm PBH} \lesssim 2.5 \times 10^{14} \rm g$ (red solid line). 
Here, we have set the free parameter $f$ in Eqs.~(\ref{ion}) and~(\ref{heat}) as $f=1$. 
The line is truncated due to the lifetime range of PBH considered in this paper. 
For comparison, the upper limits from WMAP3 (black dot-dashed line) 
and EGB (green dotted line) are also shown~\citep{lz_decay,carr}.
}
\label{fig:bp}
\end{figure}

Using Eqs.~(\ref{eq2}) and (\ref{eq:beta}), 
one can obtain the upper limits on the initial mass fraction of PBH for different masses, 
the results are shown in Fig.~\ref{fig:bp}. 
For comparison, the upper limits on $\beta(M_{\rm PBH})$ from WMAP3 
and EGB are also shown~\citep{lz_decay,carr}. 
For the mass $M_{\rm PBH} \sim 10^{14} \rm g$, the initial mass fraction is 
$\beta(M_{\rm PBH}) \sim 10^{-28}$, and it is about 2 orders of magnitude stronger 
compared with that obtained from WMAP3 data~\citep{carr}. 
There are many other 
observations which have been used to constrain $\beta(M_{\rm PBH})$ for different masses of PBHs. 
For the mass range considered by us, the limits on $\beta(M_{\rm PBH})$ 
are mainly from the extragalactic antiprotons, extragalactic neutrinos and 
the extragalactic photon background (EGB)~\citep{COMPTEL,EGRET,fermilat,carr}. 
Among of them, the most stringent constraints are from the EGB. 
With the non-observation of the excess in EGB,~\citet{carr} found a upper limit 
$\beta(M_{\rm PBH}) \sim 10^{-25}$ for $M_{\rm PBH} \sim 10^{14} \rm g$. 
On the other hand, utilizing the relevant results obtained from the WMAP3 data~\citep{lz_decay}, 
~\citet{carr} found a upper limit 
$\beta(M_{\rm PBH}) \sim 10^{-26}$ for $M_{\rm PBH} \sim 10^{14} \rm g$. 
The future 21cm surveys, such as the Square Kilometer Array, could give 
a much better upper limit, 
e.g. $\beta(M_{\rm PBH}) \sim 10^{-29}$ for $M_{\rm PBH} \sim 10^{14} \rm g$, 
if the foreground can be removed totally~\citep{mack_21}. 

It should be noticed that for the results shown in Fig.~\ref{fig:bp} we have set 
the parameter $f=1$, which means that all of the energies of electrons 
and positrons emitted by PBHs deposit into the IGM. In general, 
$f$ is a function of redshift, and it is also different for different particles. 
For more detailed discussions, one can refer to e.g.~\citet{energy_function}.

\section{constraints on the primordial curvature perturabation}
After obtaining the limits on $\beta(M_{\rm PBH})$, 
in this section we briefly review how the initial mass fraction of PBHs relates to the primordial curvature perturbation, 
and we get the limits on the power spectrum of the primordial curvature perturbation. 
 
At the end of inflation, PBHs can be formed at the scales which have left the 
horizon. 
The scales which never leave the horizon during inflation can also form 
PBHs~\citep{forma_pbhs_1}. For more detailed discussions about the formation of PBHs 
one can refer to e.g.~\citet{pbhs_review}. 
The primordial density perturbation could be gaussion or non-gaussion~\citep{forma_pbhs_2}. 
In this work, we considered the gaussion perturbations. 
According to the Press-Schechter theory~\citep{ps}, for the gaussion perturbations, 
the initial mass fraction of PBHs can be written as 

\beqa
\beta(M_{\rm PBH}) = 2\frac{M_\mathrm{PBH}}{M_\mathrm{H}}\int^{1}_{0.3}P(\delta_{H}(R))d\delta_{H}(R),
\label{beta_sec3}
\eeqa 
where $M_\mathrm{PBH} = f_{\rm M}M_{\rm H}$, $M_{\rm H}$ is the horizon mass, $f_{\rm M}=(1/3)^{1.5}$ is 
the fraction of the horizon mass which can form PBHs~\citep{Josan:2009}. 
$\delta_{\rm H}(R)$ is the smoothed density contrast at 
horizon crossing, where $R=(aH)^{-1}$. 
$P(\delta_{\rm H}(R))$ is the probability distribution of the smoothed density contrast 
with the gaussian perturbations at horizon crossing, 

\beqa
P(\delta_{\rm H}(R))=\frac{1}{\sqrt{2\pi}\delta_{\rm H}(R)}\mathrm{exp}\left(-\frac{\delta^2_{\rm H}(R)}{2\sigma^{2}_{\rm H}(R)}\right), 
\label{eq6}
\eeqa 
where $\sigma(R)$ is the mass variance. Then the initial mass fraction of PBHs can be written as

\beqa
\beta(M_{\rm PBH}) = &&\frac{2f_{M}}{\sqrt{2\pi}\sigma_{\rm H}(R)}
 \times  \nonumber \\  
&&\int^{1.0}_{0.3}\mathrm{exp}\left(-\frac{\delta^2_{\rm H}(R)}{2\sigma^2_{\rm H}(R)}\right)
d\delta_{\rm H}(R),
\label{eq4}
\eeqa
The mass variance $\sigma(R)$ is related to the power spectrum of density perturbations, $\mathcal{P_{\delta}}(k,t)$, 
as following form, 

\beqa
\sigma^{2}(R) = \int^{\infty}_{0}W^2(kR)\mathcal{P_{\delta}}(k)\frac{dk}{k},
\label{sigma}
\eeqa
where $W(kR)$ is the Fourier transform of the window function used to smooth the density contrast. 
The power 
spectrum of primordial curvature perturbation, 
$\mathcal{P}_{\mathcal{R}}(k)$, 
is related to the power spectrum 
of primordial density perturbation as~\citep{Josan:2009} 

\beqa
\mathcal{P_{\delta}}(k) = \frac{16}{3}\left(\frac{k}{aH}\right)^2j^2_{1}(k/\sqrt{3}aH)\mathcal{P}_{\mathcal{R}}(k),
\label{Pk}
\eeqa
Substituting Eq.(\ref{Pk}) into Eq.(\ref{sigma}) 
and setting $R=(aH)^{-1}$, the mass variance is written as

\begin{equation}
\sigma^2_{\rm H}(R) = \frac{16}{3}\int^{\infty}_{0}(kR)^2j_{1}^2(kR/\sqrt{3})\mathrm{exp}(-k^2R^2)\mathcal{P}_\mathcal{R}(k)\frac{dk}{k},
\label{eq5}
\end{equation}

The integral result of Eq.~(\ref{eq5}) is dominated in the scales 
$k \sim 1/R$. Following~\citet{Josan:2009} we use the 
form of $\mathcal{P}_\mathcal{R}(k)$ 
which is valid for general slow-roll inflation models as~\citep{Kohri,Leach} 

\beqa
\mathcal{P}_\mathcal{R}(k) = \mathcal{P}_\mathcal{R}(k_0)\left(\frac{k}{k_0}\right)^{n(k_0)-1},
\eeqa
where $n(k_0)=1$, and the changes of $n(k_0)$ in the ranges being allowed by the present 
observations would effect our final results slightly~\citep{josan,Josan:2009}.

\begin{figure}
\epsfig{file=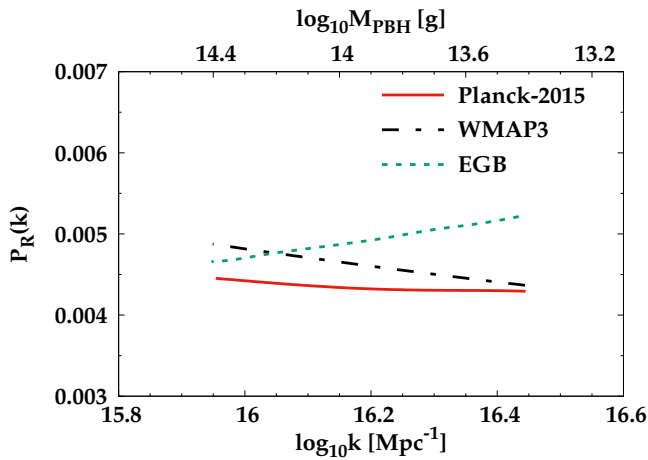,width=0.5\textwidth}
\caption{The upper limits (95\% C.L.) 
on the power spectrum of primordial curvature perturbation $\mathcal{P}_\mathcal{R}(k)$ for 
the scale range of $8.9 \times 10^{15}\ \mathrm{Mpc^{-1}} \lesssim k \lesssim 2.8 \times 10^{16}\ \mathrm{Mpc^{-1}}$ (red solid line). 
Corresponding to the Fig.~\ref{fig:bp}, the line is truncated due to the 
lifetime range of PBH considered in this paper.
For comparison, the upper limits from WMAP3 (black dotted-dashed line) and EGB (green dotted line) are also shown~\citep{lz_decay,carr}, 
which are obtained by converting the upper limits shown in Fig.~\ref{fig:bp}.
}
\label{final}
\end{figure}

Using Eq.~(\ref{eq4}), the constraints on the initial mass fraction 
of PBHs, $\beta(M_{\rm PBH})$, can be used to get the constraints 
on the mass variance, $\sigma(R)$. 
Then using Eq.~(\ref{eq5}) one can obtain the constraints on the power spectrum 
of primordial curvature perturbation, $\mathcal{P}_\mathcal{R}(k)$. 
The final results are shown in Fig.~\ref{final}. 
For the scales corresponding to the lifetime (or mass) range considered in this paper, 
$8.9 \times 10^{15} \ \mathrm{Mpc^{-1}} \lesssim k \lesssim 2.8 \times 10^{16}\  \mathrm{Mpc^{-1}}$, 
the limits do not change nearly with a value of $\mathcal{P}_\mathcal{R}(k) \sim 0.0045$. 
$\mathcal{P}_\mathcal{R}(k)$ can also be constrained 
by many other observations~\citep{Josan:2009}. 
In Fig.~\ref{final}, the upper limits on $\mathcal{P}_\mathcal{R}(k)$ 
from WMAP3 and EGB are also shown, which are obtained by converting the 
upper limits on $\beta(M_{\rm PBH})$ given in Fig.~\ref{fig:bp}. 
From Fig.~\ref{final}, it can be seen that our limits are slightly improved 
compared with that obtained from WMAP3 and EGB. 
One should be noticed that for the constraints on $\mathcal{P}_\mathcal{R}(k)$, 
we have adopted $f=1$ as done in Sec. II. 
Be similar to the constraints 
on $\beta(M_{\rm PBH})$, 
the limits on $\mathcal{P}_\mathcal{R}(k)$ are weaker for the other values of $f$ ($f<1$).

\section{conclusions}
Constraints on the power spectrum of primordial curvature perturbation 
are very important for the cosmological researches. On large 
scales, the constraints are mainly from the observations and 
researches on the CMB, Lyman-$\alpha$ and large scale structures. 
On small scales, the constraints are mainly from the researches on PBHs but these constraints 
are fairly weak. Be similar to the DM decay, PBHs have influences on 
the evolution of the IGM through the Hawking radiation. One of the results of the influences is that the anisotropies 
of the cosmic microwave background are changed. In this paper, taking into account the influences of PBHs 
on the evolution of the IGM, we used the Planck-2015 data to 
get the constraints on the initial mass fraction of PBHs and the small scale curvature perturbation. 
We focused on 
the lifetime (or mass) range of $10^{13}s \lesssim \tau_{\rm PBHs} \lesssim 10^{17}s$ 
( $2.8\times 10^{13}\mathrm{g} \lesssim M_{\mathrm{PBHs}} 
\lesssim 2.5\times 10^{14}\mathrm{g}$), which corresponds to the redshift range of 
$6\lesssim z \lesssim 1100$. We found that the optimistic upper limits are 
$4\times 10^{-29}\lesssim \beta(M_{\mathrm{PBH}}) \lesssim 5\times 10^{-28}$, 
depending on the mass of PBH. 
For the mass of $M_{\rm PBH} \sim 10^{14} \rm g$, the initial mass fraction is 
$\beta(M_{\rm PBH}) \sim 10^{-28}$, and it is about 2 (or 3) orders of magnitude stronger 
compared with that obtained from WMAP3 (or EGB). 
Using the limits on the initial mass fraction of PBHs, we got the constraints on the power spectrum of 
primordial curvature perturbation. We found that the upper limits change slightly with a value of 
$\mathcal{P}_\mathcal{R}(k) \sim 0.0045$ 
in the range of scales $8.9 \times 10^{15} \ \mathrm{Mpc^{-1}} \lesssim k \lesssim 2.8 \times 10^{16}\  \mathrm{Mpc^{-1}}$, 
which corresponds to the lifetime (or mass) range considered in this work. 
The constraints on $\mathcal{P}_\mathcal{R}(k)$ are slightly 
improvement compared with that obtained from WMAP3 and EGB.

\section{Acknowledgments}
We thank the anonymous referee and assistant editor for the very usefull suggestions and comments. 
Y. Yang thanks Xiaoyuan Huang and Yichao Li for very useful suggestions and discussions. 
Y. Yang thanks Prof. Xuelei Chen and Prof. Hongshi Zong very much. 
This work is supported in part by the National Natural Science Foundation of China 
(under Grants No.11505005, No.U1404114 and No.11373068).
\

\bibliographystyle{mnras}
\bibliography{refs}

\end{document}